# Artificial intelligence for diagnosing and predicting survival of patients with renal cell carcinoma: Retrospective multi-center study

Siteng Chen[1*], Xiyue Wang[2*], Jun Zhang[3*], Liren Jiang[4*], Ning Zhang[1], Feng Gao[4], Wei Yang[3], Jinxi Xiang[3], Sen Yang[3], Junhua Zheng[5#], Xiao Han[3#]

[1] Department of Urology, Shanghai General Hospital, Shanghai Jiao Tong University School of Medicine, Shanghai 200080, China.

[2] College of Computer Science, Sichuan University, Chengdu 610065, China.

[3] Tencent AI Lab, Shenzhen 518057, China.

[4] Department of Pathology, Shanghai General Hospital, Shanghai Jiao Tong University School of Medicine, Shanghai 200080, China

[5] Department of Urology, Renji Hospital, Shanghai Jiao Tong University School of Medicine, Shanghai 200135, China

*Equal contributors and co-first authors

#Corresponding authors:

Junhua Zheng, Department of Urology, Renji Hospital, Shanghai Jiao Tong University School of Medicine, Shanghai 200080, China. E-mail: zhengjh0471@sjtu.edu.cn. Tel: 86-021-63240090.

Xiao Han, Tencent AI Lab, Shenzhen 518057, China. E-mail: haroldhan@tencent.com. Tel: 86-075586013388.

**Abstract**

**Background:** Clear cell renal cell carcinoma (ccRCC) is the most common renal-related tumor with high heterogeneity. There is still an urgent need for novel diagnostic and prognostic biomarkers for ccRCC.

**Methods:** We proposed a weakly-supervised deep learning strategy using conventional histology of 1752 whole slide images from multiple centers. Our study was demonstrated through internal cross-validation and external validations for the deep learning-based models.

**Results:** Automatic diagnosis for ccRCC through intelligent subtyping of renal cell carcinoma was proved in this study. Our grade$_{risk}$ achieved aera the curve (AUC) of 0.840 (95% confidence interval: 0.805-0.871) in the TCGA cohort, 0.840 (0.805-0.871) in the General cohort, and 0.840 (0.805-0.871) in the CPTAC cohort for the recognition of high-grade tumor. The OS$_{risk}$ for the prediction of 5-year survival status achieved AUC of 0.784 (0.746-0.819) in the TCGA cohort, which was further verified in the independent General cohort and the CPTAC cohort, with AUC of 0.774 (0.723-0.820) and 0.702 (0.632-0.765), respectively. Cox regression analysis indicated that grade$_{risk}$, OS$_{risk}$, tumor grade, and tumor stage were found to be independent prognostic factors, which were further incorporated into the competing-risk nomogram (CRN). Kaplan-Meier survival analyses further illustrated that our CRN could significantly distinguish patients with high survival risk, with hazard ratio of 5.664 (3.893-8.239, $p < 0.0001$) in the TCGA cohort, 35.740 (5.889-216.900, $p < 0.0001$) in the General cohort and 6.107 (1.815 to 20.540, $p < 0.0001$) in the CPTAC cohort. Comparison analyses conformed that our CRN outperformed current prognosis indicators in the prediction of survival status, with higher concordance index for clinical prognosis.

**Conclusion:** Deep learning-based pathology signature could be used for the diagnosis and prognosis prediction for ccRCC, which might provide intelligent advice to improve the process of individualized treatment.

**Background**

Renal-related malignant tumor is one of the most common malignant tumors worldwide. In 2015, the incidence rate of renal cancer arrived at 66.8 per 100,000 in China [1]. In the United States, renal cancer is estimated to have 76,080 new cases and 13,780 associated deaths in 2021 [2]. Among all of the solid lesion within the kidney, renal cell carcinoma (RCC) is the most common renal-related tumor, accounting for approximately 90% of all kidney malignancies. According to cellular morphological characteristics, RCC is mainly divided into three subtypes, including clear cell RCC (ccRCC), papillary RCC (pRCC), and chromophobe RCC (ChRCC) [3]. However, some reports for ccRCC by experienced pathologists might miss essential elements and lack appropriate information associated prognosis [4]. In addition, traditional diagnosis of ccRCC by pathologist is still time-consuming and labor-intensive.

Recently, the pathology ecosystem has been gradually challenged by the emergence of digital pathology, which has also catalyzed the popularization and application of computer-aided diagnosis. Deep learning, which can be performed as a representation-learning method, has been successful used in medical image analysis with massive amounts of well-annotated data. For gigapixel whole-slide images (WSIs), they are usually annotated at the slide-level without considering the detailed internal cellular composition. Due to the gigapixel size and heterogeneity tissue distribution within the WSI, usually only a tiny region could be matched with the corresponding slide-level label, which makes the WSI-level classification problem a weakly supervised learning scenario [5-7].

Some studies have preliminarily demonstrated the utility of weakly-supervised deep learning in kidney segmentation and tumor classification from single center [8, 9]. It is also widely recognized that nuclear grading of cancer cell could act as a prognostic factor for patients with ccRCC [10]. However, traditional assessment with manual observation of nuclear grading may lead to inconsistent judgement between pathologists [11]. Moreover, there are still limitations in current TNM staging system, resulting in an urgent need for novel diagnostic and prognostic biomarkers.

In this study, we developed deep learning strategies to conduct automatic diagnosis, tumor grading, and prognosis prediction for RCC based on multi-source patient cohorts. Our study suggested that deep learning-based pathology signature could be used for the diagnosis and prognosis prediction for RCC, which might provide intelligent advice to improve the process of individualized treatment.

**Materials and methods**

**Patient cohorts and data sources**

In this study, three independent patient cohorts from different sources, including Shanghai General Hospital, Clinical Proteomic Tumor Analysis Consortium (CPTAC, https://www.cancerimagingarchive.net) [12, 13], and the Cancer Genome Atlas (TCGA, https://portal.gdc.cancer.gov) [12] were included. All included patients should meet the following selection criteria: (i) pathologically diagnosed as RCC without other types of malignant tumors; (ii) with corresponding clinical and pathological information (ground-truth label in slide-level); (iii) with access to corresponding H&E slides or their scanned WSIs.

The General cohort recruited 401 patients from the Shanghai General Hospital, who underwent partial or radical nephrectomy and were pathologically diagnosed as RCC from January 2012 to September 2019. In addition, 26 patients with renal oncocytoma were also enrolled from Shanghai General Hospital for differential diagnosis analysis. The hematoxylin-eosin staining (H&E)-stained slides were scanned with Leica Aperio AT2 scanners at 20× equivalent magnification. Furthermore, 820 patients from the TCGA cohort with diagnostic WSIs, and 195 patients from the CPTAC cohort with WSIs, who met the inclusion criteria mentioned above, were also included. The basic clinical characteristics of the included patients in this study were shown in Supplementary Table S1. Another 400 H&E-stained WSIs of RCC from the Pathology AI Platform (PAIP, wisepaip.org/paip), which were manually annotated in pixel-level by the pathologist of the Seoul National University Hospital were collected for the training and internal validation of RCC segmentation (PAIP cohort).

**Hybrid neural network for RCC segmentation**

We proposed a hybrid network for RCC segmentation as shown in Figure 1, which combined a U-net and a multi-task learning strategy to capture representative features by sharing the encoder in three task-specific branches. There are two pixel-level RCC region segmentation branches with shared five decoder layers, which are trained using the whole dataset (RCC $_{whole-seg}$) and the positive data (RCC $_{tumor-seg}$), respectively. The RCC $_{whole-seg}$ branch aims to learn distinctive features for normal and cancerous regions, which helps to reduce the rate of false positivity. The RCC $_{tumor-seg}$ branch tagetes for the more robust features to recognize tumor regions. The third classification branch (RCC $_{class}$) adopts the idea of deep supervision, which acts as an auxiliary binary classifier to determine whether an input image is positive or not.

SE-ResNeXt-50 is employed as our encoder, which is a combination of ResNeXt architecture and squeeze-and-excitation (SE) module. The ResNeXt aggregates parallel residual structures to build a wider and complex network, and the SE applies the channel attention to enhance informative feature extraction. For each decoder layer, the trainable transposed convolution operator (TransConv) is used to up-sample feature maps. These features are further connected with features in its corresponding encoder layer via skip-concatenations to preserve the consistently spatial information. Then, two convolutional layers with a batch normalization (BN) layer, a rectified linear unit (ReLU), and a selective kernel module (SKM) [14] are utilized to adaptively learn the multi-scale features. The output of the decoder is a segmentation map (256 ×256 ×1), indicating the probability of being tumor. The loss function is the combination of segmentation loss (i.e., Dice) and classification loss (binary cross-entropy) in the three branches, which was defined in our previous report [15].

**Attention-based weakly-supervised deep learning strategy**
As illustrated in Figure 2, our classification procedure can be classified into two parts: patch-level feature extraction based on self-supervised learning (SSL) and WSI-level feature aggregation based on a deep attention mechanism. For the detailed procedure, we first crop the entire WSI into small image patches (1024*1024) and then feed these patches into the pretrained SSL feature extractor [32]. to obtain a descriptive 1024-dimensional feature vector for each patch. These obtained patch-level feature vectors are assembled by deep-attention-based pooling to represent the WSI-level feature information. Referring to the attention weight of each patch, the attention pooling would average the representative features of a WSI for prediction. Two fully-connected layers following rectified linear unit (ReLU) are used to conduct WSI-level classification. In the interpretability analysis process, heatmaps, which generated by the attention weights, are used to visualize the possible disease regions that are highlighted in warm colors.

**Binary variable definition**
For patients with ccRCC, binary classification (high or low) was used for the prediction of nuclear grade, in which high grade was defined as the collection of grade III and grade IV. The overall survival (OS) status at 5-year follow up was used for the training of the prognosis-related models.

**Statistical analysis**
Continuous variants among different groups were analyzed and compared by analysis of variance. The Dice score was set as the evaluation metrics evaluate the performance of our hybrid network in tumor segmentation. Survival analysis was performed via Kaplan–Meier (KM) curve with hazard ratio (HR) and 95% confidence interval (CI) to compare different OS outcomes. We also carried out receiver operating characteristic curve (ROC) analysis with area under curve (AUC) to evaluate the accuracy of the prediction models.

**Results**
**Pixel-wise segmentation of RCC in the PAIP cohort**
A total of 400 H&E-stained WSIs of RCC with pixel-level manual annotations from the PAIP cohort was randomly divided at the patient level for the training (80%) and internal validation (20%) of the tumor segmentation model. Evaluated by five-fold cross-validation in the PAIP cohort, our hybrid network achieved a mean Dice score of 0.796 in the cross-validation cohort, exhibiting satisfactory performance of our novel hybrid architecture for pixel-wise RCC segmentation from of H&E-stained WSIs, which was independent of a classification model.

As shown in Figure 3A, our segmentation model could accuracy distinguished tumor region, which included attentional regions with high diagnostic importance while ignoring regions of low diagnostic relevance. Our hybrid network was generally capable of delineating the boundary between tumor and normal renal tissue with smooth mask (green). Insight into the magnifying representation of histopathology images indicated that the marked regions principally included tissues with dyskaryosis and structure invasion, which were also the typically morphology recognized by pathologists in clinical practices, while the normal renal tissue and other tissue, i.e. fiber texture and stroma tissue, were not included in the attentional region (Figure 3B).

**Intelligent diagnosis of RCC in the external validation cohort**
We further verified our model in an external validation cohort, which combined 928 WSIs (RCC slide: 916, normal renal slide: 12) from the TCGA cohort and 757 WSIs (RCC slide: 504, normal renal slide: 253) from the CPTAC cohort. Since the validation dataset comprised both tumor and normal images without pixel-level annotations, which was more in accordance with the clinical practices, we assigned a probability value of RCC to a test image if the area of segmentation occupied more than 5% of the WSI after removing the white space. Based on the strategy, the AUC for distinguishing RCC from normal renal tissue achieved 0.977 (95% CI: 0.969-0.984, Figure 3C) in the in an external validation cohort, which borne comparison with an experienced pathologist.

Further subgroup analysis based on the subtypes of RCC revealed that our diagnosis model could diagnosis clear cell RCC, papillary RCC, and chromophobe RCC from normal renal tissues, with AUCs of 0.987 (0.979-0.993), 0.939 (0.913-0.960), and 0.984 (0.961-0.995), respectively (Figure S1), which indicating the robust generalization performance of our model when applied to different scenarios.

**Interpretability and whole-slide attention visualization**
Readable interpretability of deep learning-based clinical models plays important role in further clinical applications [16]. To gain insight into the potential interpretability of our model, we visualized the learned feature space in two dimensions to generate pixel-level heatmaps. As shown in the Figure S1 (right column), the most attended regions recognized by our model were considered to be highly associated with RCC. Areas with red color of the heatmap represented the regions with predicted RCC tissues. The pixel-level visualization by our model presented the spatial distributions of diverse tissues, which also helped to provide human-in-the-loop interaction to optimize the current diagnostic processes.

**Differential diagnosis of RCC from renal oncocytoma**
Renal oncocytoma was one of the most common benign tumors in renal, which had several features that overlapped with RCC with a preponderance of granular cytoplasm [17]. Misconceptions could be reviewed out in clinical practice due to the Review out spectrum of eosinophilic renal neoplasms. Therefore, we further explored whether our hybrid network could be used for the differential diagnosis of RCC from renal oncocytoma in clinical practices. As shown in the Figure S2, our diagnosis model exhibited excellent performance in the differential diagnosis of RCC, which achieved an AUC of 0.951 (0.922-0.972), a sensitivity of 0.821 (0.772-0.862), and a specificity of

0.962 (0.804-0.999).

**Intelligent subtyping of RCC through deep learning**

Clinical outcomes differ remarkably among patients with different subtypes of RCC, and ccRCC causes worse prognosis than pRCC and ChRCC [18]. Since the identification of different subtypes plays a vital role in clinical practices, we proposed a novel neural network for the intelligent subtyping of RCC based on a weakly-supervised deep learning strategy.

As shown in Figure 4A, our subtyping model performed well in the subtype prediction of RCC, with an average AUC of 0.990 (95% CI: 0.981-0.996) in distinguishing ccRCC from pRCC and ChRCC in the TCGA cross-validation cohort, which could be used for the automatic diagnosis of ccRCC. The classification accuracy was further verified in the General cohort, with AUC of 0.970 (0.957-0.980, Figure 4B). Visualization of the subtyping model revealed that our diagnosis model could recognize the tumor regions with transparent and gelatinous material, which contributed to the accurate diagnosis ccRCC (Figure 4C).

**Recognition of high-grade tumor through deep learning**

The prognostic value of the nuclear grading has been widely recognized for patients with ccRCC [3, 19]. Therefore, we further applied the weakly-supervised learning strategy to predict high-grade tumors for the grade-classification of ccRCC. The model was trained and cross-verified from the TCGA cohort and was based on the hypothesis that some microscopic features associated with high-grade tumors could be identified and integrated to calculate the $\text{grade}_{\text{risk}}$ for the automatic recognition of high-grade ccRCC. As shown in Figure S3A, our $\text{grade}_{\text{risk}}$ achieved an average AUC of 0.840 (0.805-0.871) in the TCGA cross-validation cohort for distinguishing high-grade tumors, which was further verified in the independent General cohort and the CPTAC cohort, with AUC of 0.840 (0.805-0.871, Figure S3C) and 0.840 (0.805-0.871, Figure S3E), respectively. Comparation analyses indicated that the $\text{grade}_{\text{risk}}$ distributed differently among patients with different tumor grades (Figure S3B, D, F), which further confirmed the potential for clinical practice.

**Intelligent risk quantitation for five-year survival follow-up**

Clear cell RCC accounts for most of the adverse prognosis related to renal malignancy. Therefore, it is of great importance to accurately predict the 5-year OS status and quantify the survival risk for patients in clinical follow-up. Based on the weakly-supervised learning strategy, we assembled patch-level feature vectors with attention weight to conduct WSI-level classification of 5-year OS status. The survival risk for 5-year follow-up ($\text{OS}_{\text{risk}}$) was then calculated based on the prediction possibility. As illustrated in Figure S4A, our $\text{OS}_{\text{risk}}$ achieved an average AUC of 0.784 (0.746-0.819) in the TCGA cross-validation cohort for identifying patient with adverse clinical outcome in 5-year follow-up, which was further verified in the independent General cohort and the CPTAC cohort, with AUC of 0.774 (0.723-0.820, Figure S4D) and 0.702 (0.632-0.765, Figure S4G), respectively. Further comparation analyses revealed that our $\text{OS}_{\text{risk}}$ distributed differently among patients with different tumor grades (Figure S4B, D, G) and different tumor grades (Figure S4C, E, H). Patients with higher tumor grades or stages seemed to have higher $\text{OS}_{\text{risk}}$, which was consistent with the clinical observations that patients with higher tumor grades/stages might suffer from more survival risk and less likely to get a five-year survival follow-up.

**Development of the competing-risk nomogram**

Integration of multiple biomarkers might improve predictive value over single-scale counterpart [20, 21]. We had proved that deep learning-based pathology signatures, including the grade$_{risk}$ and the OS$_{risk}$, were significantly associated with high-grade tumor and 5-year survival status. Therefore, we next to explore whether our deep learning-based pathology signatures could cooperate with traditional clinicopathological characteristics to improve the prognosis prediction for clinical practice.

We firstly carried out cox regression analysis to identify prognostic indicators. As shown in Figure 5A, the grade$_{risk}$, the OS$_{risk}$, tumor grade, and tumor stage were found to be independent prognostic factors for patient with ccRCC. These four factors were further incorporated into the construction of the competing-risk nomogram (CRN, Figure 5B). ROC analysis revealed that when the cut-off value was set as 103, our CRN achieved the best performance in predicting the OS status in 5-year follow-up, with the highest AUC of 0.825 (0.789-0.858), specificity of 0.902, and sensitivity of 0.637. With the same cut-off value, patients in the TCGA cohort were classified into the worse group or the favorable group. Kaplan-Meier survival analyses further illustrated that our CRN could significantly distinguish patients with high survival risk (Figure 6A), with HR of 5.664 (95% CI 3.893-8.239, p < 0.0001). Verification of our CRN in the independent General cohort (Figure 6B) and the CPTAC cohort (Figure 6C) further confirmed the robust prognostic power, with HR of 35.740 (5.889-216.900, p < 0.0001) and 6.107 (1.815 to 20.540, p < 0.0001), respectively.

**Comparison with current prognosis indicators**

To further identify the superiority of our CRN in prognosis prediction of ccRCC, we compared the CRN with current prognosis indicators through multiple indexes, including AUCs for 5-year, 3-year, 1-year OS status and the concordance index (C-index). As shown in Table 1, our CRN outperformed current prognosis indicators in the prediction of 5-year, 3-year, 1-year OS status. The CRN achieved the highest C-index value from 0.770 to 0.846, which overmatched current prognosis indicators. In addition, CRN also achieved higher AUCs in the prediction of 5-year, 3-year, 1-year OS status when compared to the comprehensive clinicopathology feature (Figure 6D-L).

**Discussion**

Traditional visual inspection of pathological images can be distinguished by the nuclear shape, size, nucleolus, and chromatin features. For renal carcinoma with high tumor heterogeneity, traditional microscope vision may miss a lot of important information. Furthermore, the shortage of pathologists has aggravated the presence of overwork in pathology. In the United States, the absolute pathologist workforce had decreased from 2007 to 2017, which resulted in the increase of the diagnostic workload by about 42% [22]. There is still an urgent need to develop novel technologies to prevent potential diagnostic error from traditional pathology.

The application of deep neural networks in digital pathology has greatly catalyzed the intelligent analysis of pathological image, otherwise it cannot be analyzed by human-based image interrogation [23]. DL with CNN demonstrates consummate performance in multiple prediction task from pathological WSI, including tissue segmentation [24], cancer diagnosis [25], cancer prognosis [26], and mutation prediction [27]. Excellent performance of DL has also been reported in displaying distinct immunogenomic landscape and potential response to immunotherapy [28, 29].

Based on the full landscapes of WSIs, a deep CNN was reported to identify different subtypes of RCC [30]. A histopathology image classifier could also distinguish TFE3 Xp11.2 translocation RCC from ccRCC, which contributed to overcome the difficulties that could not be easily solved in traditional analysis through naked eye [31]. Benefiting from the increasing number of image datasets, AI-based approaches are now defining integrated and clinically classification of RCC. However, most of the AI-based models were trained from comparatively small samples, without sufficient additional validation.

Currently, the diagnosis reports of WSIs are usually at the global level (slide-level). However, the slide-level labels are often associated with tiny/small regions from the gigapixel WSI, which turns the WSI-level classification problem into a weakly-supervised learning scene (i.e., inexact supervision). To tackle this problem, we performed the multiple-instance learning to achieve WSI-level classification in view of the entire information from the slide. Since the gigapixel WSIs could not be directly feed into network, we segmented the WSI into non-overlapped patches with 1024*1024 pixels at 20× magnification. All patches extracted from the same slide were then identified as the instances of a specific WSI. It is noted that WSI were labeled in slide-level annotations of tumor region, and thus, these extracted patches have no annotations.

Through the application of CNN, we proposed an end-to-end neural network for the diagnosis and prognosis prediction of RCC. With a WSI input, the network could achieve automatic and rapid diagnosis, grading, and survival prediction for the patient. To our knowledge, this is the largest cohort used in our neural network for the classification of ccRCC using H&E-stained WSIs. The subtype identification performed well in the internal and external validation cohorts, with the matched sensitivity and specificity of an experienced pathologist, but substantial workload had been saved through our network. In addition, we also provided convincing predictions survival status, which might facilitate clinical decision-making but could not be provided through traditional pathology.

In this study, we proposed a data-efficient weakly-supervised learning strategy to address the annotation lack problems in the field of histopathological images. Recently, a clustering-constrained-attention multiple-instance learning framework (CLAM) was also proposed to improve the weakly-supervised learning [16], which was further applied to AI-based assessment of tumor origins [25]. There are two major differences between this study and ours. First, CLAM adopts pretrained model based on natural images as the feature extractors. The huge domain shift between natural and histopathological images may decrease the model generalization. We encode the semantic content of each patch using our previous pretrained feature extractor on large-scale and diverse histopathological images in an unsupervised manner. Second, we conduct a multi-task learning for comprehensive RCC stratifications, including cancer/nuclei subtyping and prognosis/mutation prediction, whereas CLAM performs a single task for cancer subtype classifications.

Several strengths could be found in this study. Firstly, adequate WSIs from three independent patient cohorts were recruited for training and testing the deep neural network, which improved the generalization performance of our models. Secondly, with only a WSI input, the weakly-supervised network makes it possible for automatic and rapid classification for ccRCC. Thirdly, based on the importance scores of sub-regions in the WSI, an interpretable probability map can be generated to point out the diagnostically relevant regions for pathologists, making it more practical to clinical practice.

There are also some limitations waiting for solution in our study. Firstly, part of the images analyzed in this study were acquired for public databases, which might be affected by the potential population bias. Secondly, batch effect might be involved in this analysis since different H&E-staining protocols might be performed among different patient cohorts. Thirdly, this is a retrospective study, which might need further validations in prospective clinical studies.

**Conclusions**

In summary, we proposed a weakly-supervised deep learning strategy for the diagnosis and prognosis prediction of RCC with interpretable probability. Using conventional histology, our method could achieve automatic diagnosis, tumor grading, and prognosis prediction for patients with ccRCC, thereby providing intelligent advice to improve the process of individualized treatment.

**Acknowledgements**
We appreciate the partial image data from Clinical Proteomic Tumor Analysis Consortium, the Cancer Genome Atlas, and the Cancer Imaging Archive used in this study.

**Authors' contributions**
JHZ and XH conceptualized and supervised the study. STC, XYW, and JZ performed data curation, formal analysis, investigation, visualization, and writing original draft. LRJ, NZ, FG, WY, SY and JXX performed data curation, and validation. All authors involved manuscript editing and manuscript review.

**Funding**
This study was supported by the National Natural Science Foundation of China (81972393). The funders had no role in the design of the study and collection, analysis, and interpretation of data and in writing the manuscript.

**Availability of data and materials**
Primary data are available from Atlas (https://portal.gdc.cancer.gov/) and the Clinical Proteomic Tumor Analysis Consortium (https://www.cancerimagingarchive.net/). Other private data could only be reasonably requested from the corresponding author according to the Research Ethics Committee.

**Declarations**
**Ethics approval and consent to participate**
Our study was approved by the Research Ethics Committee of Shanghai General. Consents were acquired form the participates.

**Consent for publication**
Not applicable.

**Competing interests**
The authors declare that they have no competing interests.

**References**

[1] Chen W, Zheng R, Baade PD, Zhang S, Zeng H, Bray F, et al. Cancer statistics in China, 2015. CA Cancer J Clin. 2016;66:115-32.
[2] Siegel RL, Miller KD, Fuchs HE, Jemal A. Cancer Statistics, 2021. CA Cancer J Clin. 2021;71:7-33.
[3] Ljungberg B, Bensalah K, Canfield S, Dabestani S, Hofmann F, Hora M, et al. EAU guidelines on renal cell carcinoma: 2014 update. Eur Urol. 2015;67:913-24.
[4] Shuch B, Pantuck AJ, Pouliot F, Finley DS, Said JW, Belldegrun AS, et al. Quality of pathological reporting for renal cell cancer: implications for systemic therapy, prognostication and surveillance. BJU Int. 2011;108:343-8.
[5] Chen PC, Gadepalli K, MacDonald R, Liu Y, Kadowaki S, Nagpal K, et al. An augmented reality microscope with real-time artificial intelligence integration for cancer diagnosis. Nat Med. 2019;25:1453-7.
[6] Ehteshami Bejnordi B, Veta M, Johannes van Diest P, van Ginneken B, Karssemeijer N, Litjens G, et al. Diagnostic Assessment of Deep Learning Algorithms for Detection of Lymph Node Metastases in Women With Breast Cancer. Jama. 2017;318:2199-210.
[7] Nagpal K, Foote D, Liu Y, Chen PC, Wulczyn E, Tan F, et al. Development and validation of a deep learning algorithm for improving Gleason scoring of prostate cancer. NPJ Digit Med. 2019;2:48.
[8] Han S, Hwang SI, Lee HJ. The Classification of Renal Cancer in 3-Phase CT Images Using a Deep Learning Method. J Digit Imaging. 2019;32:638-43.
[9] Tanaka T, Huang Y, Marukawa Y, Tsuboi Y, Masaoka Y, Kojima K, et al. Differentiation of Small (≤ 4 cm) Renal Masses on Multiphase Contrast-Enhanced CT by Deep Learning. AJR Am J Roentgenol. 2020;214:605-12.
[10] Delahunt B, Eble JN, Egevad L, Samaratunga H. Grading of renal cell carcinoma. Histopathology. 2019;74:4-17.
[11] Holdbrook DA, Singh M, Choudhury Y, Kalaw EM, Koh V, Tan HS, et al. Automated Renal Cancer Grading Using Nuclear Pleomorphic Patterns. JCO Clin Cancer Inform. 2018;2:1-12.
[12] Clark K, Vendt B, Smith K, Freymann J, Kirby J, Koppel P, et al. The Cancer Imaging Archive (TCIA): maintaining and operating a public information repository. J Digit Imaging. 2013;26:1045-57.
[13] Clark DJ, Dhanasekaran SM, Petralia F, Pan J, Song X, Hu Y, et al. Integrated Proteogenomic Characterization of Clear Cell Renal Cell Carcinoma. Cell. 2019;179:964-83.e31.
[14] Li X, Wang W, Hu X, Yang J. Selective Kernel Networks.  2019 IEEE/CVF Conference on Computer Vision and Pattern Recognition (CVPR)2020.
[15] Wang X, Fang Y, Yang S, Zhu D, Wang M, Zhang J, et al. A hybrid network for automatic hepatocellular carcinoma segmentation in H&E-stained whole slide images. Med Image Anal. 2021;68:101914.
[16] Lu MY, Williamson DFK, Chen TY, Chen RJ, Barbieri M, Mahmood F. Data-efficient and weakly supervised computational pathology on whole-slide images. Nat Biomed Eng. 2021;5:555-70.
[17] Amin MB, Crotty TB, Tickoo SK, Farrow GM. Renal oncocytoma: a reappraisal of morphologic features with clinicopathologic findings in 80 cases. Am J Surg Pathol. 1997;21:1-12.
[18] Patard JJ, Leray E, Rioux-Leclercq N, Cindolo L, Ficarra V, Zisman A, et al. Prognostic value of histologic subtypes in renal cell carcinoma: a multicenter experience. J Clin Oncol. 2005;23:2763-71.
[19] Delahunt B, Cheville JC, Martignoni G, Humphrey PA, Magi-Galluzzi C, McKenney J, et al. The

International Society of Urological Pathology (ISUP) grading system for renal cell carcinoma and other prognostic parameters. Am J Surg Pathol. 2013;37:1490-504.

[20] Chen D, Liu Z, Liu W, Fu M, Jiang W, Xu S, et al. Predicting postoperative peritoneal metastasis in gastric cancer with serosal invasion using a collagen nomogram. Nat Commun. 2021;12:179.

[21] Jiang Y, Zhang Q, Hu Y, Li T, Yu J, Zhao L, et al. ImmunoScore Signature: A Prognostic and Predictive Tool in Gastric Cancer. Ann Surg. 2018;267:504-13.

[22] Metter DM, Colgan TJ, Leung ST, Timmons CF, Park JY. Trends in the US and Canadian Pathologist Workforces From 2007 to 2017. JAMA Netw Open. 2019;2:e194337.

[23] Barisoni L, Lafata KJ, Hewitt SM, Madabhushi A, Balis UGJ. Digital pathology and computational image analysis in nephropathology. Nat Rev Nephrol. 2020;16:669-85.

[24] Pantanowitz L, Quiroga-Garza GM, Bien L, Heled R, Laifenfeld D, Linhart C, et al. An artificial intelligence algorithm for prostate cancer diagnosis in whole slide images of core needle biopsies: a blinded clinical validation and deployment study. Lancet Digit Health. 2020;2:e407-e16.

[25] Lu MY, Chen TY, Williamson DFK, Zhao M, Shady M, Lipkova J, et al. AI-based pathology predicts origins for cancers of unknown primary. Nature. 2021;594:106-10.

[26] Courtiol P, Maussion C, Moarii M, Pronier E, Pilcer S, Sefta M, et al. Deep learning-based classification of mesothelioma improves prediction of patient outcome. Nat Med. 2019;25:1519-25.

[27] Coudray N, Ocampo PS, Sakellaropoulos T, Narula N, Snuderl M, Fenyö D, et al. Classification and mutation prediction from non-small cell lung cancer histopathology images using deep learning. Nat Med. 2018;24:1559-67.

[28] Xie F, Zhang J, Wang J, Reuben A, Xu W, Yi X, et al. Multifactorial Deep Learning Reveals Pan-Cancer Genomic Tumor Clusters with Distinct Immunogenomic Landscape and Response to Immunotherapy. Clin Cancer Res. 2020;26:2908-20.

[29] Sealfon RSG, Mariani LH, Kretzler M, Troyanskaya OG. Machine learning, the kidney, and genotype-phenotype analysis. Kidney Int. 2020;97:1141-9.

[30] Marostica E, Barber R, Denize T, Kohane IS, Signoretti S, Golden JA, et al. Development of a Histopathology Informatics Pipeline for Classification and Prediction of Clinical Outcomes in Subtypes of Renal Cell Carcinoma. Clin Cancer Res. 2021;27:2868-78.

[31] Cheng J, Han Z, Mehra R, Shao W, Cheng M, Feng Q, et al. Computational analysis of pathological images enables a better diagnosis of TFE3 Xp11.2 translocation renal cell carcinoma. Nat Commun. 2020;11:1778.

[32] Wang X, Du Y, Yang S, et al. RetCCL: Clustering-guided contrastive learning for whole-slide image retrieval[J]. Medical Image Analysis, 2023, 83: 102645..

**Table 1** Comparison with current prognosis indicators

|  | 5-year OS status | | 3-year OS status | | 1-year OS status | | C-index |
|---|---|---|---|---|---|---|---|
|  | AUC | 95% CI | AUC | 95% CI | AUC | 95% CI |  |
| **TCGA cohort** | | | | | | | |
| Grade | 0.708 | 0.666-0.747 | 0.718 | 0.676-0.757 | 0.726 | 0.685-0.764 | 0.667 |
| Stage | 0.764 | 0.724-0.800 | 0.794 | 0.756-0.828 | 0.822 | 0.786-0.854 | 0.729 |
| Grade $_{risk}$ | 0.723 | 0.682-0.762 | 0.733 | 0.692-0.771 | 0.725 | 0.684-0.763 | 0.677 |
| OS $_{risk}$ | 0.785 | 0.747-0.820 | 0.779 | 0.740-0.814 | 0.812 | 0.775-0.845 | 0.727 |
| CRN | 0.825 | 0.789-0.858 | 0.841 | 0.806-0.871 | 0.869 | 0.837-0.897 | 0.770 |
| **General cohort** | | | | | | | |
| Grade | 0.798 | 0.748-0.841 | 0.848 | 0.802-0.886 | 0.943 | 0.910-0.966 | 0.820 |
| Stage | 0.788 | 0.738-0.833 | 0.842 | 0.796-0.881 | 0.856 | 0.812-0.893 | 0.800 |
| Grade $_{risk}$ | 0.799 | 0.750-0.843 | 0.880 | 0.839-0.915 | 0.882 | 0.841-0.916 | 0.820 |
| OS $_{risk}$ | 0.774 | 0.723-0.820 | 0.885 | 0.844-0.919 | 0.870 | 0.828-0.906 | 0.803 |
| CRN | 0.814 | 0.766-0.856 | 0.924 | 0.888-0.951 | 0.969 | 0.943-0.986 | 0.846 |
| **CPTAC cohort** | | | | | | | |
| Grade | 0.677 | 0.607-0.742 | 0.694 | 0.624-0.758 | 0.627 | 0.556-0.695 | 0.659 |
| Stage | 0.796 | 0.733-0.850 | 0.803 | 0.740-0.856 | 0.748 | 0.680-0.807 | 0.773 |
| Grade $_{risk}$ | 0.693 | 0.624-0.757 | 0.711 | 0.642-0.773 | 0.685 | 0.615-0.750 | 0.689 |
| OS $_{risk}$ | 0.702 | 0.632-0.765 | 0.708 | 0.639-0.771 | 0.679 | 0.609-0.744 | 0.684 |
| CRN | 0.803 | 0.740-0.856 | 0.809 | 0.747-0.862 | 0.754 | 0.688-0.813 | 0.780 |

# Figures

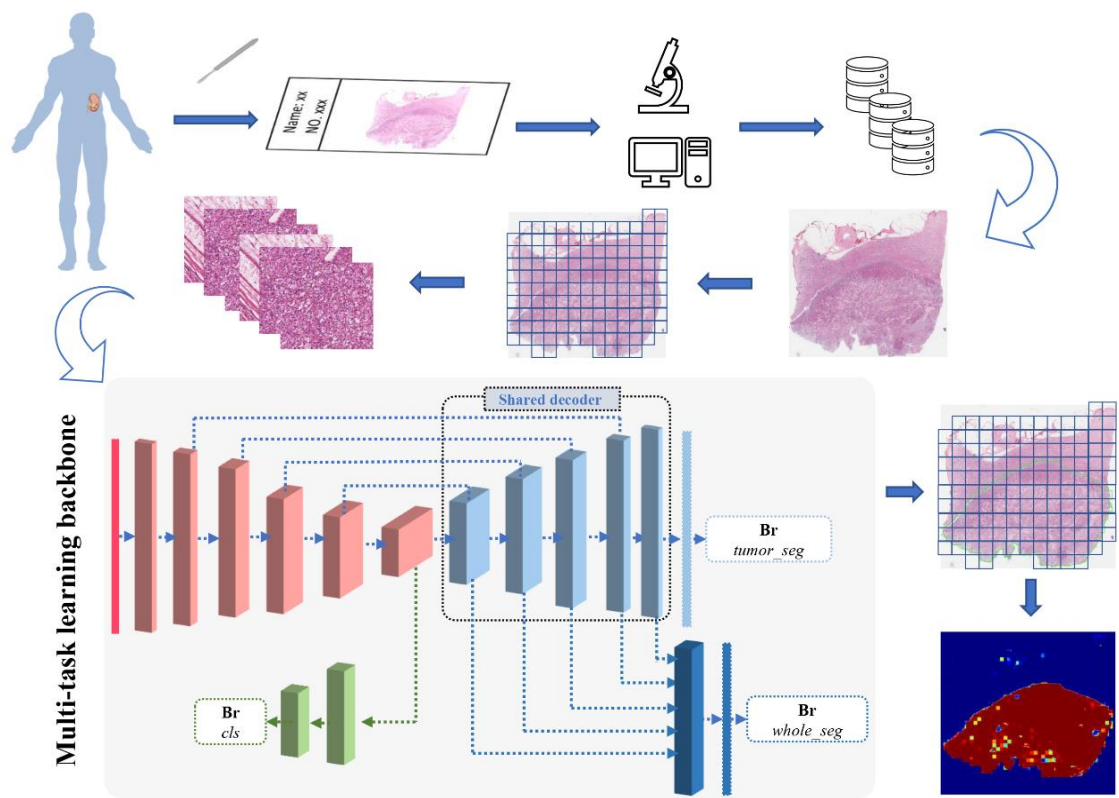

**Figure 1** The workflow and the architecture of the hybrid neural network proposed in this study. Br $_{whole\_seg}$, pixel-wise renal cell carcinoma segmentation; Br cls, auxiliary classification task.

**Figure 2** Architecture of weakly-supervised learning strategy based on multiple instance-learning scene for subtype diagnosis, grade staging, and survival analysis of renal cell carcinoma. FC, fully connected.

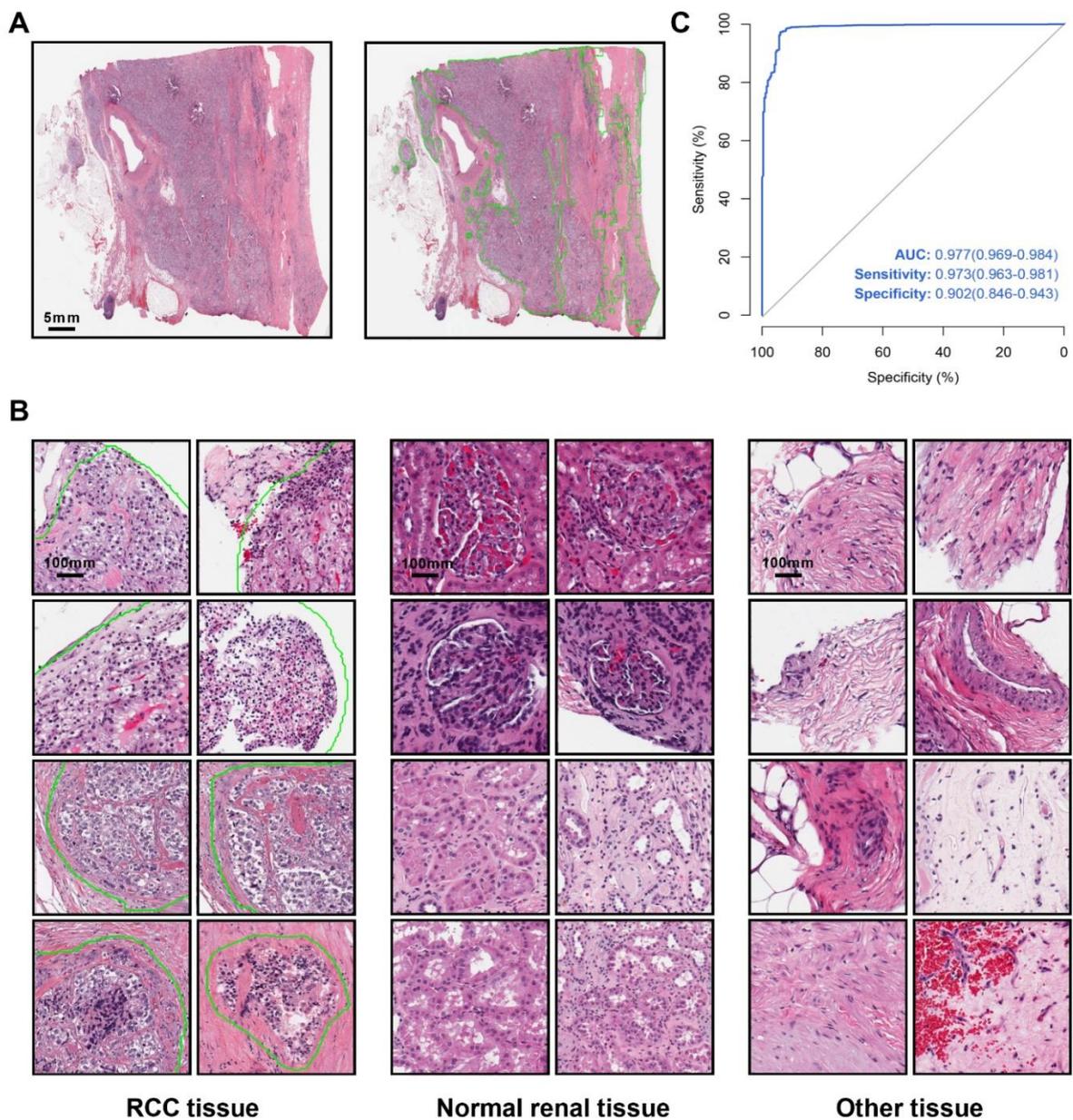

**Figure 3** Accurate segmentation of RCC for intelligent diagnosis. (A) Example RCC segmentation. Left, original slide image; Right, recognized slide image with green curve. (B) Example of segmentation on different kinds of tissue. (C) ROC curve for distinguishing RCC from normal renal tissues in the independent verification cohort. RCC, renal cell carcinoma; ROC, receiver operator characteristics; AUC, area under the curve (with 95% confidence interval).

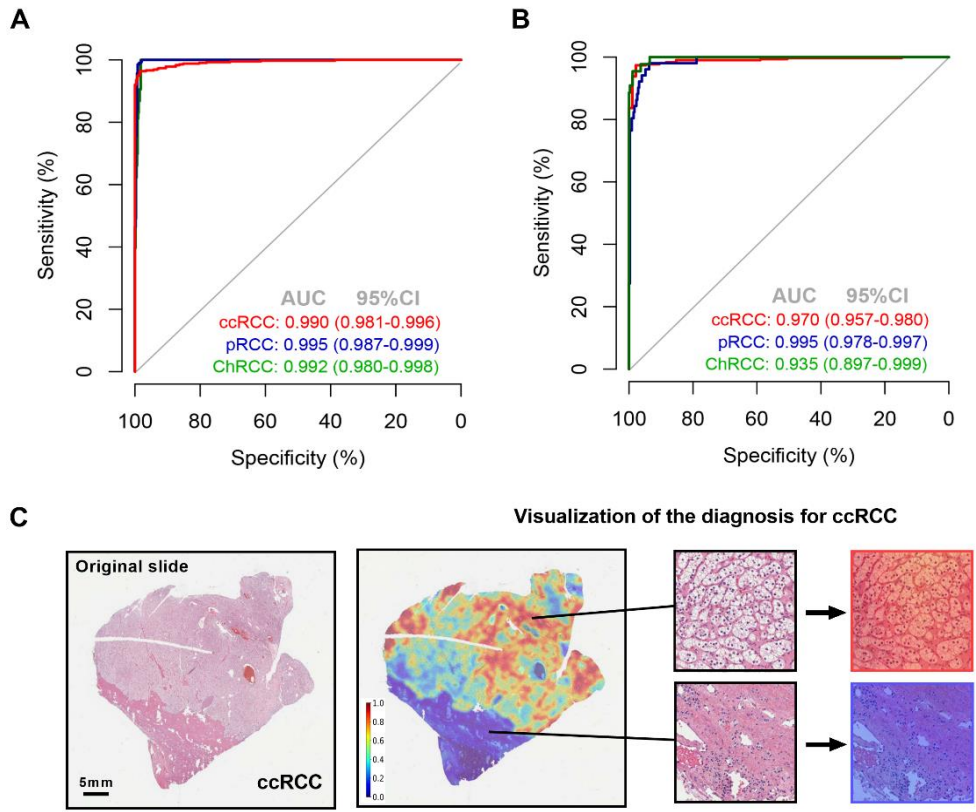

**Figure 4** Intelligent subtyping of renal cell carcinoma from weakly-supervised learning. (A, B) ROC curves for intelligent subtyping of RCC in the cross-validation cohort (The Cancer Genome Atlas cohort) and the validation cohort (General cohort), respectively. (C) Visualizations of the diagnosis for ccRCC. The detected tumor regions were shown in red. ccRCC, clear cell renal cell carcinoma; pRCC, papillary renal cell carcinoma; ChRCC, chromophobe renal cell carcinoma; ROC, receiver operator characteristics; AUC, area under the curve; CI, confidence interval.

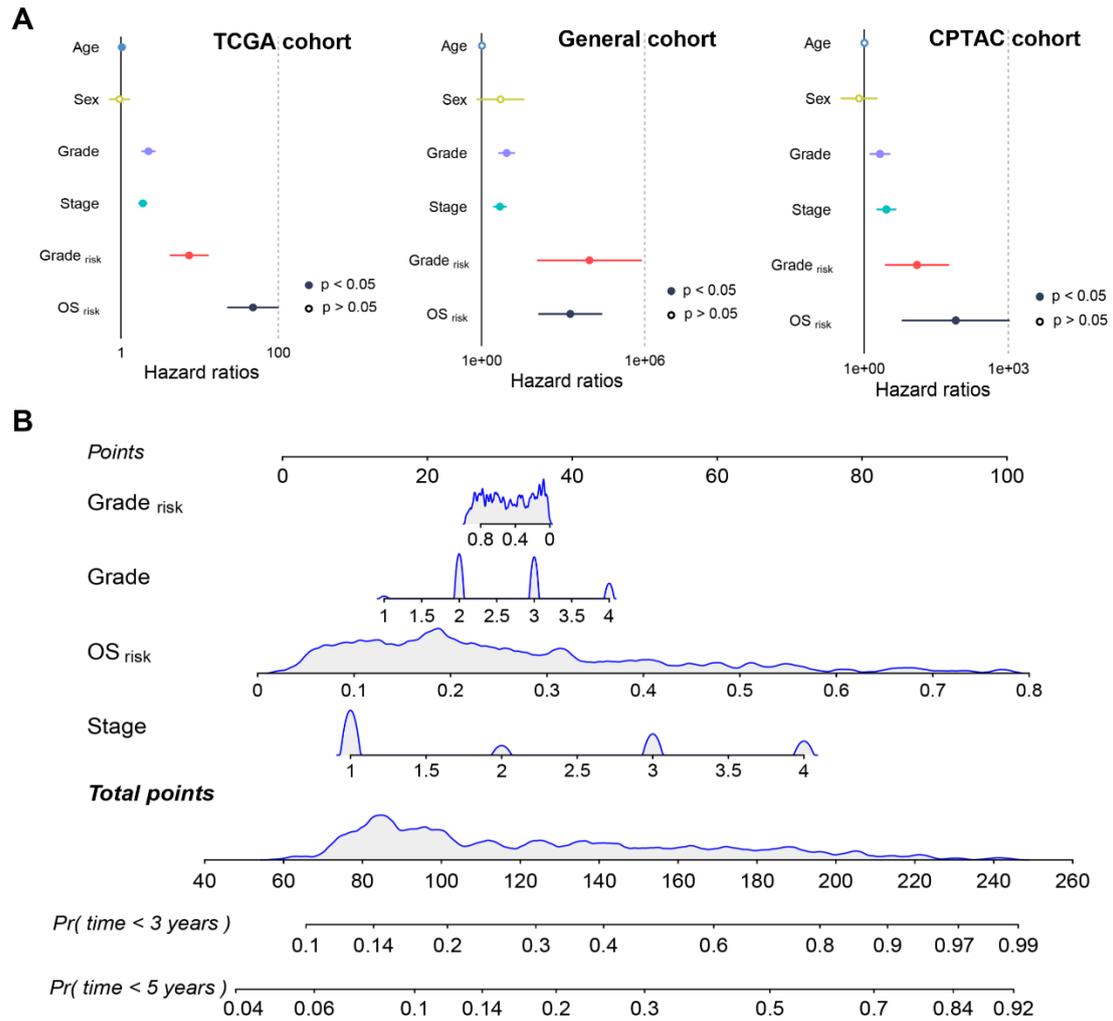

**Figure 5** Construction of the competing-risk nomogram. (A) Cox regression analyses of the deep learning-based pathology signature and clinicopathological features. (B) The competing-risk nomogram for the construction of the prognosis prediction model combining the deep learning-based pathology signature and clinicopathological features. TCGA, the Cancer Genome Atlas; CPTAC, Clinical Proteomic Tumor Analysis Consortium; OS, overall survival.

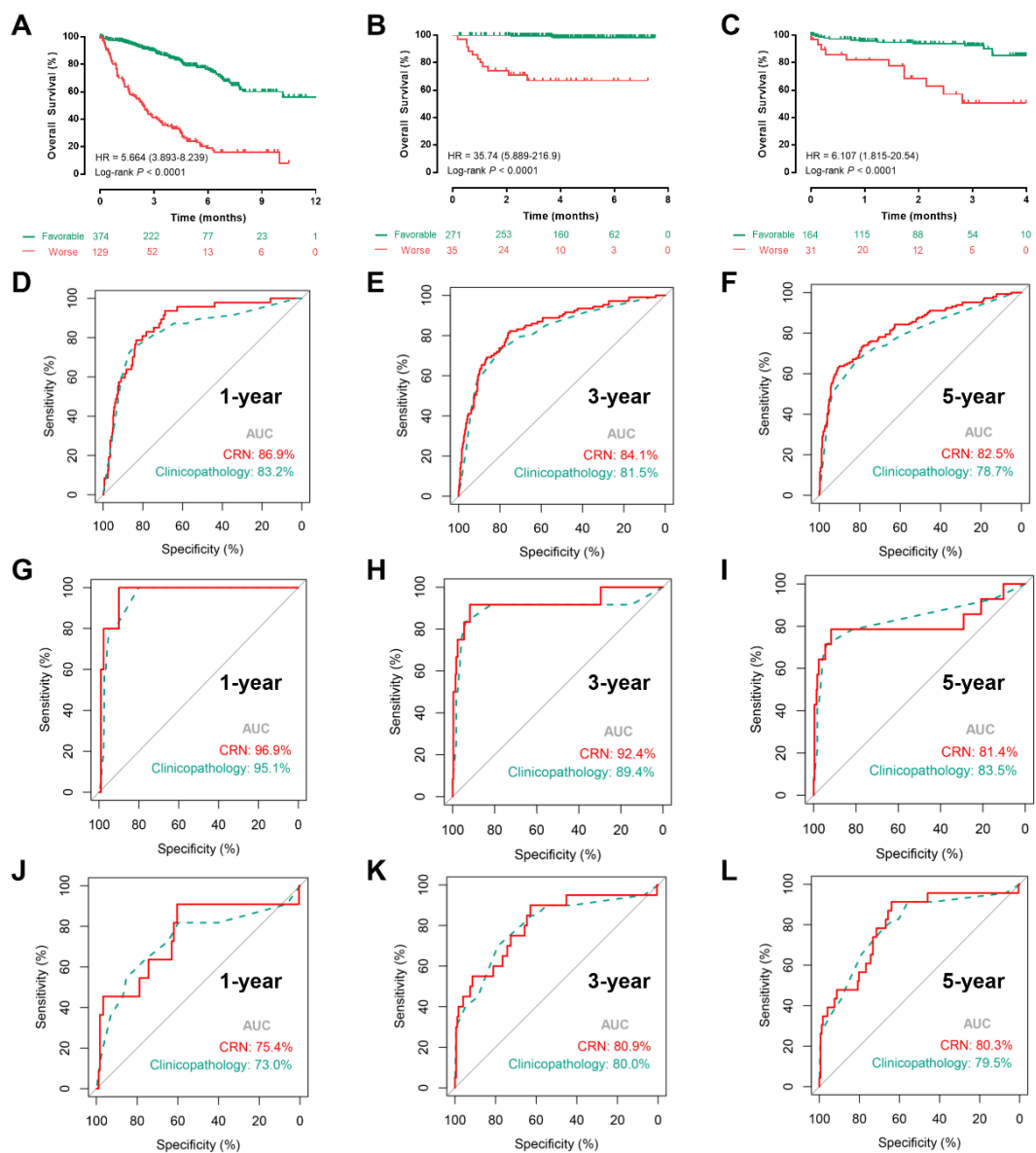

**Figure 6** Evaluations of the CRN model. (A) Kaplan-Meier survival analysis of overall survival in the Cancer Genome Atlas cohort. (B) Kaplan-Meier survival analysis of overall survival in the General cohort. (C) Kaplan-Meier survival analysis of overall survival in the Clinical Proteomic Tumor Analysis Consortium cohort. (D, E, F) ROC curves of 1-, 3-, and 5-year overall survival prediction for the CRN model and comprehensive clinicopathology features in the Cancer Genome Atlas cohort. (G, H, I) ROC curves of 1-, 3-, and 5-year overall survival prediction for the CRN model and comprehensive clinicopathology feature in the General cohort. (J, K, L) ROC curves of 1-, 3-, and 5-year overall survival prediction for the CRN model and comprehensive clinicopathology features in the Clinical Proteomic Tumor Analysis Consortium cohort. CRN, competing-risk nomogram, ROC, receiver operator characteristics; AUC, area under curve.

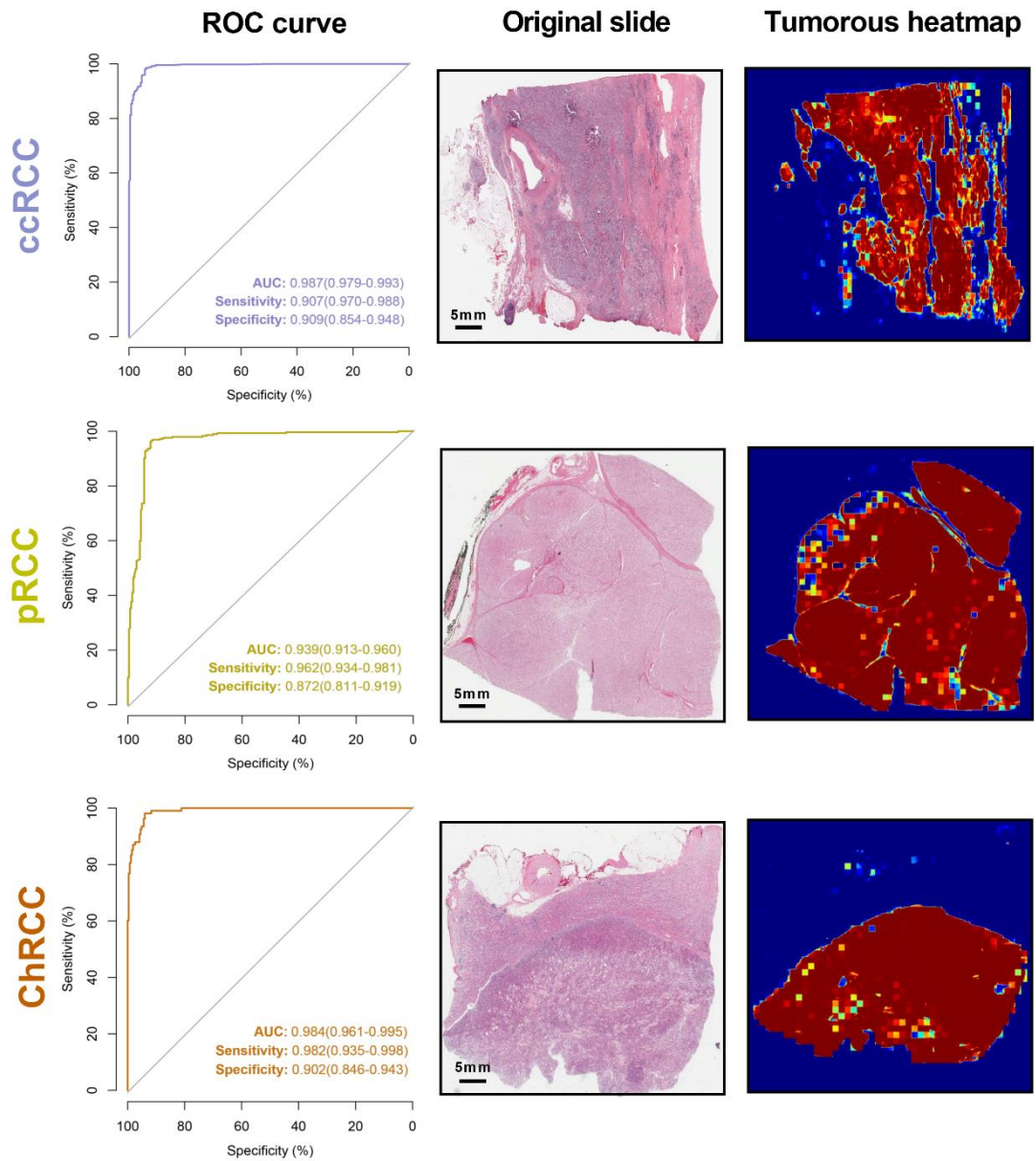

**Figure S1** Accurate diagnosis of ccRCC, pRCC, and ChRCC from normal renal tissues. Left, ROC curves for distinguishing RCC from normal renal tissues; Middle, original slide images; Right, visualization of detected tumor regions for each type of RCC; RCC, renal cell carcinoma; ccRCC, clear cell renal cell carcinoma; pRCC, papillary renal cell carcinoma; ChRCC, chromophobe renal cell carcinoma; ROC, receiver operator characteristics; AUC, area under the curve (with 95% confidence interval).

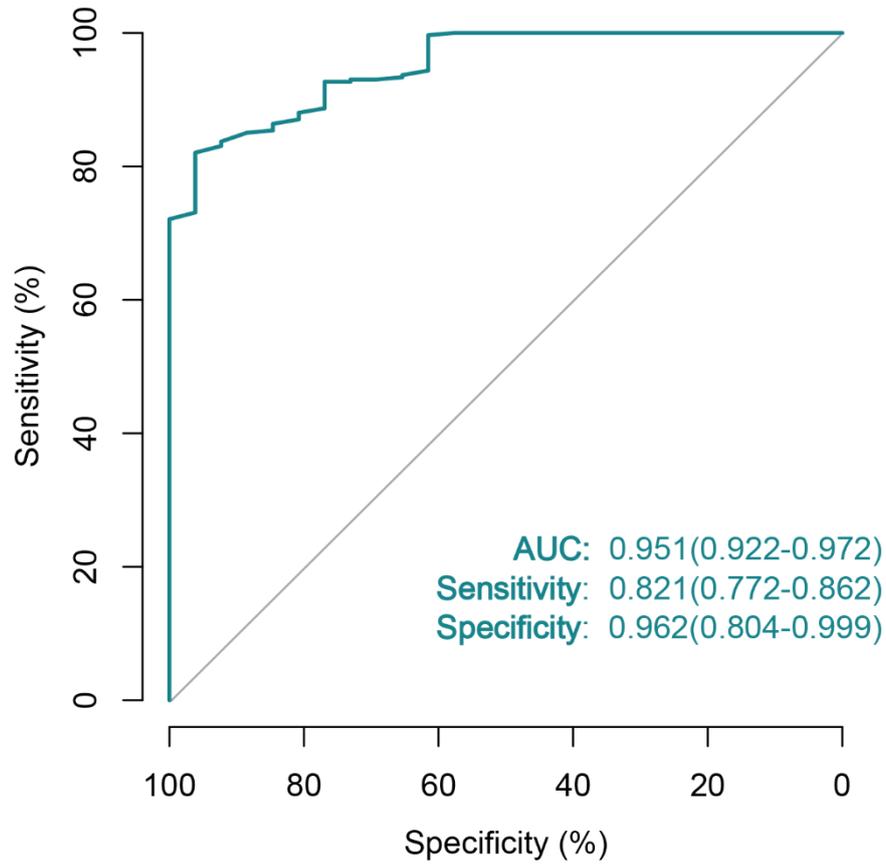

**Figure S2** Differential diagnosis of renal cell carcinoma from renal oncocytoma. AUC, area under the curve (with 95% confidence interval).

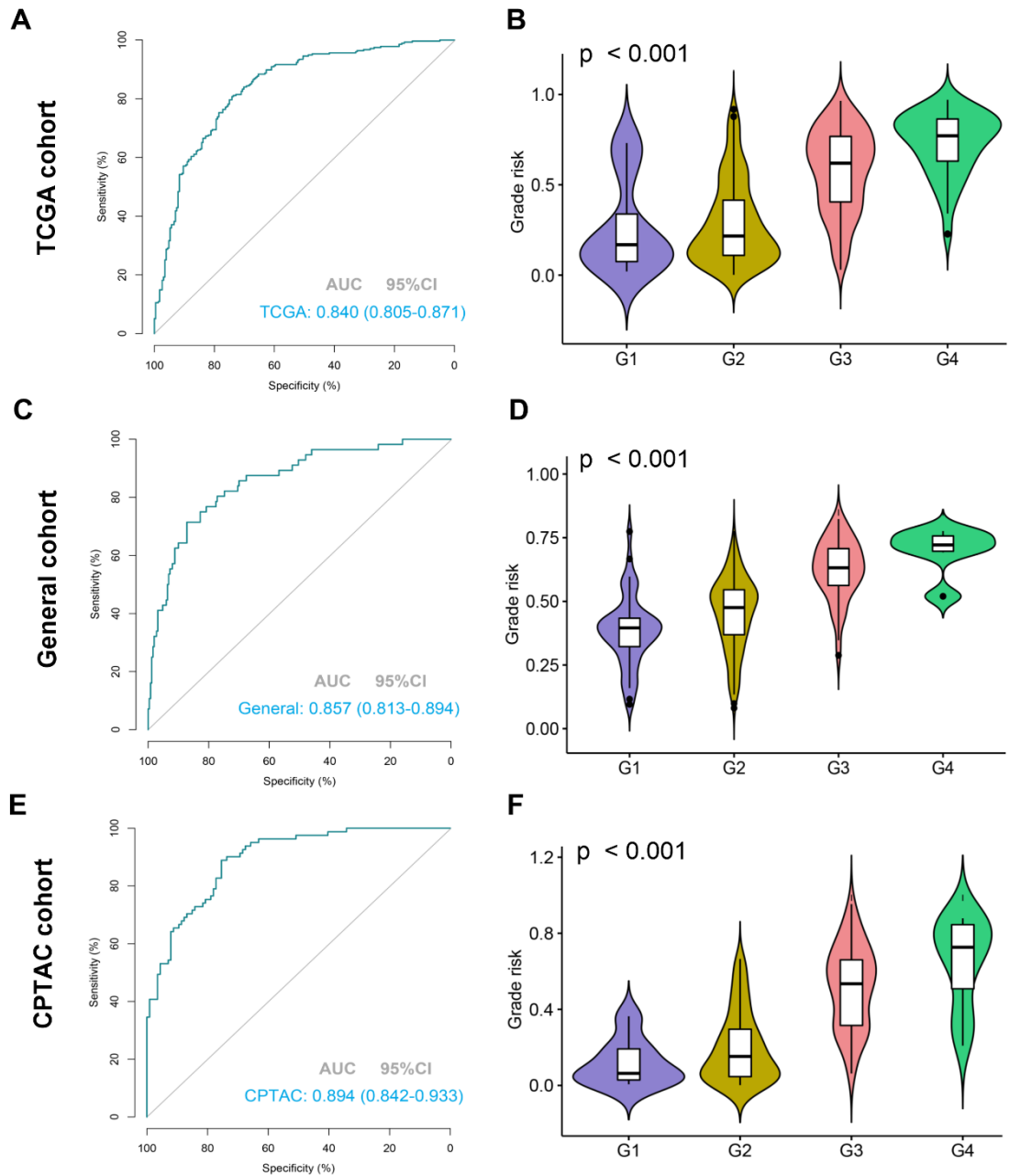

**Figure S3** Prediction of high tumor grade for patients with clear cell renal cell carcinoma. (A, C, E) ROC curves for the prediction of high tumor grade for ccRCC in the TCGA cohort, the General cohort, and the CPTAC cohort, respectively. (B, D, F) Comparisons of the grade$_{risk}$ among patients with different tumor grades in the TCGA cohort, the General cohort, and the CPTAC cohort, respectively. ROC, receiver operator characteristics; AUC, area under the curve; TCGA, the Cancer Genome Atlas; CPTAC, Clinical Proteomic Tumor Analysis Consortium; CI, confidence interval.

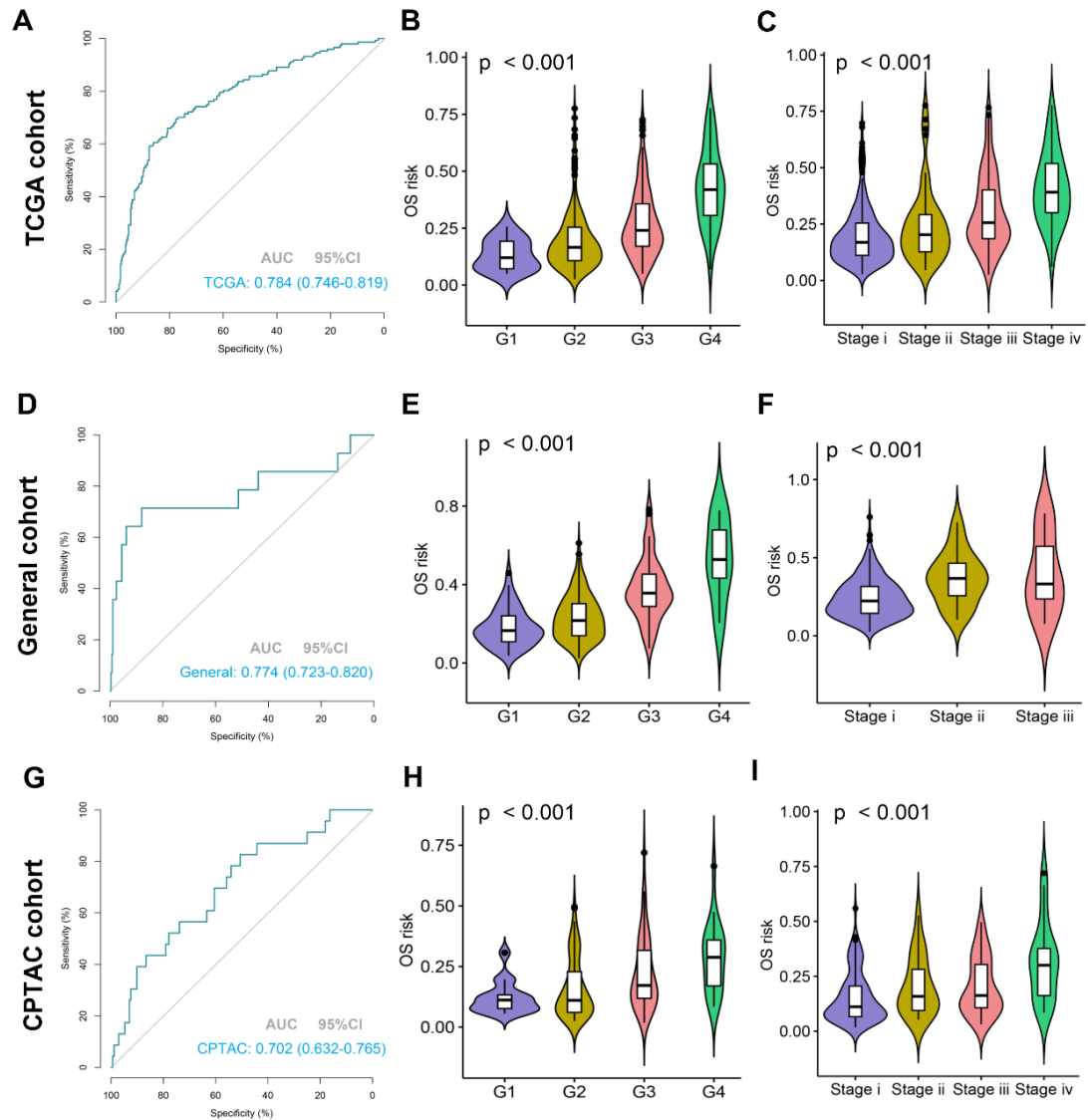

**Figure S4** Prediction of the 5-year OS status for patients with clear cell renal cell carcinoma. (A, D, G) ROC curves for the prediction of the 5-year OS status for ccRCC in the TCGA cohort, the General cohort, and the CPTAC cohort, respectively. (B, E, H) Comparations of the $OS_{risk}$ among patients with different tumor grades in the TCGA cohort, the General cohort, and the CPTAC cohort, respectively. (C, F, I) Comparations of the $OS_{risk}$ among patients with different tumor stages in the TCGA cohort, the General cohort, and the CPTAC cohort, respectively. OS, overall survival; ROC, receiver operator characteristics; AUC, area under the curve; TCGA, the Cancer Genome Atlas; CPTAC, Clinical Proteomic Tumor Analysis Consortium; CI, confidence interval.

# Supplemental Table

**Table S1** Basic clinical characteristics of patients recruited for this study.

|  | **General Cohort (401)** | **TCGA Cohort (820)** | **CPTAC Cohort (195)** |
|---|---|---|---|
| **Age(years)** | | | |
| ≥65 | 139(34.7%) | 263(32.1%) | 78(40.0%) |
| < 65 | 262(65.3%) | 557(67.9%) | 117(60.0%) |
| **Sex** | | | |
| Male | 288(71.8%) | 548(66.8%) | 138(70.8%) |
| Female | 113(28.2%) | 272(33.2%) | 57(29.2%) |
| **Stage** | | | |
| i | 362(90.3%) | 434(52.9%) | 100(51.3%) |
| ii | 25(6.2%) | 105(12.8%) | 20(10.3%) |
| iii | 14(3.5%) | 182(22.2%) | 54(27.7%) |
| iv | 0 | 99(12.1%) | 21(10.7%) |
| **WSI** | 401 | 847 | 195 |
| **Subtype** | | | |
| **ChRCC** | 44(11.0%) | 65(7.9%) | / |
| **pRCC** | 51(12.7%) | 244(29.8%) | / |
| **ccRCC** | 306(76.3%) | 511(62.3%) | 195(100%) |
| **Nuclear grade** | | | |
| High (iii/iv) | 56(18.3%) | 275(53.8%) | 81(41.5%) |
| Low (i/ii) | 250(81.7%) | 228(44.6%) | 114(58.5%) |
| Unknown | 0 | 8(1.6%) | 0 |
| **Status** | | | |
| Dead | 14(5.6%) | 170(33.3%) | 23(11.8%) |
| Alive | 292(95.4%) | 333(65.2%) | 172(88.2%) |
| Unknown | 0 | 8(1.5%) | 0 |